# Navigating the Sociotechnical Imaginaries of Brazilian Tech Workers

Kenzo Soares Seto

## Introduction

Current research on the relation between work and digital technologies often narrows its focus to specific interests such as platform labor, gig workers and content creators experiences, or the increased social power of Global North's platforms. While these studies offer valuable insights, they frequently overlook a critical group: the tech workers who design and develop these digital systems (Dorschel, 2020a; Browne et al., 2024). Data scientists, developers, and software engineers play a fundamental role in shaping the systems that permeate our lives and understanding their perspectives is essential for a comprehensive analysis of how technology interacts with society.

Tech workers occupy a unique position at the intersection of labor and technology. Although often subordinated within large corporations, they possess significant bargaining power due to the high demand for their specialized skills (Selling & Strimling, 2023). Their reflections and experiences provide crucial insights into the agency they exercise over the systems they create, as they influence not only technical design but also the ethical, social, and political dimensions embedded in algorithms and platforms (Noble, 2018). Despite this, there is a notable gap in the literature addressing contemporary tech workers' perspectives in the digital economy, especially within the context of the Global South (Seto, 2024a).

As the social impacts of digital technologies become more apparent it becomes imperative to investigate the sociotechnical imaginaries of the workers who develop these technologies. Sociotechnical imaginaries, as defined by Mager and Katzenbach (2021), are collectively constructed visions of the future, shaped by shared social values. These imaginaries guide and legitimize technological development and governance, influencing how societies conceive, develop, and utilize new tools and systems.

Studying the imaginaries of tech workers in the Global South is crucial to challenge "data universalism," the hegemonic narrative that overlooks geographical boundaries and social contexts, imposing a one-size-fits-all approach to technology and data (Milan and Treré, 2019). By exploring the unique perspectives of tech workers in countries like Brazil, we can uncover alternative technological imaginaries that challenge dominant frameworks and envision technologies more attuned to local needs, values, and modes of existence.

This chapter aims to fill this gap by examining the sociotechnical imaginaries of Brazilian tech workers. Drawing from my fieldwork involving developers, data scientists, and software engineers in the country, it explores how these professionals perceive the relationship between their work and society. By focusing on their concerns and visions for the future, this research contributes to a more nuanced understanding of how digital technologies can be developed and governed to promote diversity, equity, and local sovereignty. It highlights the need for methodological approaches that consider their realities and the ethical challenges regarding workers' safety and privacy. And through their voices, we gain insight into how these workers navigate the tensions between global technological advancements and local challenges.

In the next section, I explore the concepts of sociotechnical imaginaries and tech workers, providing a concise review of the key trends in the literature and establishing the theoretical framework for this study. The subsequent section details the methodologies used in tech worker studies, emphasizing ethical considerations and the importance of incorporating diverse perspectives. Building on this foundation, I then present the findings from my fieldwork, discussing Brazilian tech workers' imaginaries.

## Tech Workers' Sociotechnical Imaginaries

Sociotechnical imaginaries are shared visions of desirable futures, collectively constructed and institutionally stabilized, shaped by common understandings of social order. Rooted in interpretive social theory, their study highlights the collective imagination's power to create shared systems of meaning that underpin the production of social reality (Mager & Katzenbach, 2021).

In the context of technology, studying imaginaries challenges the popular notion that innovations arise from the creativity and entrepreneurship of individual genius. Instead, social practices and institutions influence which technical developments are prioritized or neglected (Jasanoff and Kim, 2009). Therefore, a key feature of sociotechnical imaginaries is their multiplicity, traversed by controversies among social actors like corporations, governments, and civil society, each expressing distinct visions of technology's role (Mager & Katzenbach, 2021). These imaginaries shape not only the design of technology, but also who benefits from it (Jasanoff, 2015), which is critical to understanding the emergence and challenge of hegemonic regimes of technology ownership, governance, and appropriation. In this sense, the perspectives of tech workers are fundamental. As part of a working class view of technology, they have a particular knowledge of the development of technology and how the balance of power between them and their employers affects it.

The concept of "tech workers" has boundaries subject to theoretical and political controversies. Originating in the 1950s in the field of information systems (IS), which describes professionals who design and manage information structures (Dorschel, 2022a), the category is broadly defined in two ways: sectorally and occupationally. Sectorally, tech workers are all those employed by companies that produce information and communication technologies (ICT), while occupationally, tech workers include only those in knowledge-intensive roles related to coding in various industries (Rothstein, 2022).

This strict definition of the concept can unreflectively obscure the broader relationship between technology and labor, ignoring that all human labor has a technological aspect (Seto, 2024b). For example, focusing only on those who develop software can obscure the relevance of labor to the IT industry, such as hardware assembly and data curation by microworkers (Irani, 2019). Thus, some authors, such as Fuchs (2013) and Rothstein (2022), as well as labor organizations, such as the Tech Worker Coalition, seek to expand the definition of tech workers to include all workers employed by tech companies, seeing this theoretical-political operation as the most useful for building working class solidarity (Rothstein, 2022).

Conversely, Dorschel (2020a) argues for distinguishing tech workers from hardware and microtask workers, noting that tech workers generally have advanced education, specialized knowledge, relative autonomy, and better conditions than most of the working class; conflating them has analytical inaccuracies. Therefore, we define tech workers as those with expertise in the design and development of algorithmic systems, digital databases and Artificial Intelligence (AI) systems using programming languages (Zuboff, 1994; Miller & Coldicutt, 2019; Roy, 2021; Dorschel, 2022a; 2022b; Selling & Strimling, 2023). Moreover, while self-identification as a "tech worker" is complex and evolving, it forms a transnational community based on a collective sense of belonging, knowledge sharing, and shared institutions that are strongly linked to the act of coding (Tarnoff & Weigel, 2020, Beltrán, 2023).

Research on tech workers has traditionally focused on Silicon Valley. In a pioneering effort, Barbrook and Cameron (1995) characterized the dominant political imaginary there as the "California ideology," that sought to replace political processes with free-market models and technological innovation to solve social problems, what Morozov (2018) calls "Techno-solutionism". This led to the displacement of collective organizing inherited from the counterculture, replaced by entrepreneurship and meritocratic competition in the 1980s (Turner, 2006), with English-Lueck (2011) suggesting that the tech industry's meritocratic culture was so strong that it undermined birth-assigned identities like gender. Conversely, Noble (2018) argues that the fact that the tech worker community is predominantly white, male, and privileged is a reason for the persistence of racist and patriarchal biases in Silicon Valley systems, reproducing epistemologies rooted in colonial mechanisms of control.

However, Selling and Strimling (2023), Niebler (2023), as well as Rothstein (2022), note that recent unionization movements reflect a shift among tech workers toward more leftist worldviews, challenging the hegemony of California ideology. Similarly, Dorschel (2022b) identifies a "post-neoliberal subjectivity" emerging among them, characterized by a critique of economic inequality and a concern for diversity and inclusion. This concern express as growing reflections about social implications of their work, such as bias in AI systems (Miller and Coldicutt, 2019; Tarnoff & Weigel, 2020; Browne et al., 2024), and critical perspectives about the ethical gaps in the corporative discourse of "Technology for Good" (Powell et al., 222).

Studies have also focused on understanding racialized, queer, and gendered tech workers' experiences and imaginaries, from Indigenous (Nakamura, 2014) and Black women's work (Shetterly, 2016) in the genesis of the digital industry to how racialized North American tech workers confront racism since the IBM Black Workers Alliance from the sixty's (Niebler, 2023) to the present (Alfrey, 2014; Franklin, 2020; Chow, 2023). In the feminist and queer fields, social risks and biases in algorithms and AI are discussed from feminist approaches to the views of tech workers (Browne et al., 2024) and from the perspectives of LGBTQIA+ tech workers (Myles et al. (2023), while debating how diversity, equality, and inclusion (DEI) agendas affect the culture and imaginaries of the IT industry (Alfrey and Twine, 2017; Alfrey, 2022; Browne et al., 2024).

Beyond the global North, research on tech workers is concentrated in Asia. In China, state vigilance makes online platforms, such as GitHub, crucial for tech workers' anonymous discussions of structural sexism (Liu, 2023) and labor rights (Li, 2019, Niebler, 2023), while their concerns about AI risks are less intense than their western counterparts (Di, 2020). The emerging Chinese 'maker' culture is also challenging western intellectual property norms, with tech workers hacking patents to rework the notion of 'Made in China' and 'digital technologies with Chinese characteristics' around local community solutions (Lindtner, 2014).

In India, Irani (2019) notes that tech workers face both precarity and privilege, torn between their will to contribute to their country's development and market pressures. Despite their white-collar status, they endure job insecurity and long hours, leading to tentative union links (Roy, 2021) in the context of a growing domestic market linked to digital sovereignty perspectives (Kumar and Thussu, 2023). There is also a diasporic dimension to the Indian tech community, where their precarious status as immigrants in the US (Bhatt, 2018) or Europe (Amrute, 2016) is in tension with their aspirations to be part of an Indian national renaissance through technological development.

In Latin America, Beltrán (2023) examines how Mexican tech workers use hackathons and "migrahacks" to address issues like U.S. immigration and Mexican society's problems, combining latin identity, hacker culture,entrepreneurship and feminist movements. In Brazil, women's tech communities focus on networking and gender inequality awareness (Paz, 2015; Frade, 2021), while LGBTQIA+ tech workers debates remote work as a safe space, especially for non-binary individuals (Souza Santos et al., 2023).

Despite these studies, research on the political imaginaries of tech workers remains primarily focused on the Global North (Daub, 2020). To change this reality, it's crucial to examine the socio-technical imaginaries of tech workers worldwide, especially in underrepresented regions such as Latin America.

# Methodological approaches to studying the imaginaries of tech workers

Exploring the sociotechnical imaginaries of tech workers—how their values, culture, and beliefs influence the design and development of technology—requires a multifaceted methodological approach. One effective starting point is the use of surveys to obtain a general overview of the community: for example, Miller and Coldicutt (2019) conducted a survey of 2,234 British tech workers and found that 59% of respondents had witnessed decisions in AI development that they believed were dangerous to society.

Research on tech workers has also favored qualitative methods, from in-depth and semi-structured interviews (Didi, 2020; Dorschel, 2022b; Powell et al., 2022; Chow, 2023; Browne et al., 2024) to ethnographic fieldwork (Ustek-Spilda et al., 2019; Irani, 2019; Powell et al., 2022; Beltrán, 2023). There are many opportunities for ethnographic work: from long-term immersions in IT offices (Armute, 2016; Irani, 2019; Ustek-Spilda et al., 2019), to activist spaces such as hackathons (Beltráns, 2023), to broader conferences such as industry meetups (Powell et al., 2022). It is also possible to develop comparative case studies, as Rothstein (2022) has done with workplaces in the United States and Germany.

Netnography is another valuable method, especially in contexts where open expression and physical encounters can be risky, such as in authoritarian regimes. By examining social media posts, open letters, and manifestos, researchers can access expressions of tech workers' sociotechnical imaginaries (Li, 2019; Liu, 2023). This method can also extend to analyzing comments in code repositories or inside the code of programs, providing insights into tech workers perspectives embedded within technical discourse.

Interviews remain one of the most widely used techniques for studying sociotechnical imaginaries and researching tech workers. To this end, building trust with participants is crucial, especially due to the risks associated with workers expressing criticism and insider information about their industry, which are exacerbated by the common practice of tech workers signing non-disclosure agreements (NDAs).

**Kenzo Soares Seto (2025).Navigating the Sociotechnical Imaginaries of Brazilian Tech Workers. In Ergin Bulut, Julie Chen, Rafael Grohmann, and Kylie Jarrett (eds.). SAGE Handbook of Digital Labour. SAGE: 314-324.**

In my research with Brazilian tech workers, I used the snowball sampling method described by Vinuto (2014). In this approach, the initial participants, contacted through the Alumni mailing of an IT Faculty, recommended new contacts until the research reached a saturation point where new interviews do not yield new relevant information. The referrals from trusted peers help to validate the researcher and encourage participation. While this qualitative study provides in-depth insights, the sample size of 26 participants limits the generalizability of the findings. Additionally, the reliance on snowball sampling may introduce bias, as participants are likely to refer to individuals within similar networks.

An important difference between my research and studies conducted in the Global North is the inclusion of tech workers employed in sectors beyond Big Tech companies associated with the Internet and digital platforms. While most researchers often focus on workers in companies such as Alphabet, Meta and Amazon (Tarnoff, 2020; Jaffe, 2021; Selling and Strimling, 2023, Rothstein, 2022), my research also includes tech workers in Brazil's most competitive economic sectors, where the country's IT and AI investments are therefore concentrated: oil and gas, agribusiness, and mining (Seto, 2024c). For example, I interviewed data scientists at Petrobras, Brazil's state-owned national oil company, which operates the largest supercomputer in Latin America entirely dedicated to AI development (Petrobras, 2023).

I actively sought as much demographic diversity in my sample as possible. This effort stemmed from the epistemological concern that the perspectives of racialized, LGTQIAPN+, and gendered tech workers can provide valuable insights into how the imaginaries and epistemologies of data and computational science may shift when viewed through the experiences of marginalized groups. I also sought professional diversity by interviewing tech workers developing video games, delivery and e-commerce platforms, cybersecurity solutions for fintechs, and AI solutions for Meta, Brazilian startups, and state-owned enterprises. In addition, I captured the perspectives of professionals with no experience as political activists, as well as those involved in social movements.

**Table 1**

*Participants' socio-demographic and professional characteristics*

| | | | |
|---|---|---|---|
| **Genre and sexuality** | Men | Women | LGBTQIAP+ |
| | 16 | 10 | 5 |
| **Race** | White | Afro-brazilians | Not identified |
| | 17 | 5 | 4 |
| **Age** | 26-30 | 30-36 | |
| | 19 | 7 | |
| **Work experience** | 2 to 5 years | 5 to 10 years | >10 years |
| | 7 | 17 | 2 |
| **Employment** | Private foreign company | Private national company | State-owned company |
| | 5 | 20 | 1 |

The analysis of data from tech workers has prioritized grounded theory, identifying emerging themes and patterns in the transcription of interviews (Dorschel, 2022), or other similar methods of thematic coding (Hui et al., 2023) and open coding (Chow, 2023), the same methodology used by me. A fundamental dimension of this method is to conduct interviews as open-ended conversations rather than structured questionnaires. This empowers participants to define what is relevant, potentially revealing issues that might be missed in more rigid formats, and also expresses an ethical approach by the researcher to amplify rather than silence participant voices, especially when interviewing marginalized groups.

In studies with gendered or queer tech workers, researchers have also sought feminist and queer theoretical references to reflect on the research method and analyze the data (Irani, 2019), especially Feminist science and technology studies (STS) (Browne et al., 2024). Similarly, Seto (2024a) emphasizes the importance of analyzing the imaginaries of tech workers in dialogue with epistemologies from the Global South, mobilizing, for example, Brazilian thinker Paulo Freire's theory of the knowledge of the oppressed. This approach recognizes that tech workers in the Global South operate within contexts deeply shaped by colonial histories and economic dependencies, but also with their own traditions of social thought (Seto, 2024a).

# Navigating the Sociotechnical Imaginaries of Brazilian Tech Workers

"Technology isn't just about code. We're shaping society, and that's a heavy responsibility." – Lucas[1], senior developer

"We imagine a future where technology bridges gaps, not widens them. But is that the path we're on?" – Ana, data scientist

These reflections encapsulate the concerns and hopes of Brazilian tech workers regarding the societal impact of their work. Between July and December 2023, I conducted semi-structured interviews with 26 Brazilian tech professionals using a snowball sampling method (Vinuto, 2014). Key themes emerged from the dialogue between me and the tech workers' own reflections on the relationship between their work and society, revealing their core values and concerns about the social role of digital technology. Their perspectives highlight emerging socio-technical imaginaries, their collective visions about Brazil's technological present and future.

---
[1] To protect the anonymity of respondents, all names are pseudonyms.

A recurring theme among the interviews was the issue of algorithmic bias. Such biases can stem from underlying assumptions coded into algorithms, biased training data, or design choices that reflect programmers' own prejudices (Noble, 2018). While all interviewees acknowledged these risks, many felt that the broader Brazilian tech community seldom addresses the issue. João, a senior developer, remarked: "This topic is only strong in academia. A developer hears the phrase 'racist algorithm' and thinks, 'An algorithm is mathematics; how can a mathematical operation be racist?'"

This vision highlights a gap between academic discourse and industry practice, and a lack of integration between digital humanities and computational science in universities, where most of them are educated. João summed up the tech community's skepticism about attributing political agency to technical means:

> What happens is an anthropomorphization of the algorithm, importing a human notion of bias and prejudice into AI. The algorithm doesn't comprehend color; it doesn't have the concept of race. It just multiplies matrices. If you train the model to identify potential criminals based on previous patterns, and if most prisoners are Black, it will reproduce that. The model is just optimizing it.

Many workers argued that biases stem from datasets rather than the algorithms themselves, believing that these datasets merely reflect existing societal prejudices. Pedro, a data scientist, noted: "The bias is in the database, not in the model. The inequality isn't in the technical system; it's in society. There are no algorithms that solve structural sexism; society's problems can't be attributed to technology."

However, racialized and LGBTQIA+ professionals emphasized that the normalization of biases in technological systems results from the lack of diversity within the tech community. Kim, an experienced developer, provided an example:

When Microsoft released the Kinect, it didn't work well with Black people. Not necessarily because the programmers were racist, but because they didn't have Black programmers to test it and see there was a problem. The very fact that the developer isn't actively working against bias is bias in itself.

These findings reinforce the importance of diversity and inclusion in technology development to mitigate algorithmic biases that perpetuate systemic inequalities. However, this mitigation will not be achieved through corporate DEI agendas (Alfrey, 2022; Browne et al., 2024). It will emerge from internal struggles between workers and corporate policies. In this sense, queer imaginaries are already fuelling struggles within and against digital platforms, as the values of tech workers clash with cisheteronormative corporate guidelines. One developer, Joana, hired to adapt algorithmic filters for Portuguese content moderation on a large Chinese social platform, recounted: "We were working on improving automatic filters that identify content for human curation. The biggest surprise was when we realized there was a policy in the models to censor videos of affectionate LGBTQIA+ interactions, like hugs and kisses."

Since part of the Brazilian team was LGBTQIA+, they collectively resisted directives from the Chinese headquarters, arguing contradictions with local laws and threatening to leak evidence to the press. They successfully prevented censorship of Portuguese content but remained unaware of policies for other national markets.

Tech worker's concerns about the impact of digital technologies on users also revealed a tension between corporative logic and workers' values about systems developments. José, who worked on automating shopping experiences for a national retail platform, highlighted the lack of discussion about reinforcing consumers' compulsive behavior:

> We work with neuroscience to exploit human weaknesses to the maximum. The aim of the system is to speed up decision-making. Countdown timers induce haste; designs express scarcity; user journeys minimize clicks to complete a purchase—all to trigger unthinking emotional decisions. We've never questioned how the systems we develop affect users financially, how they mobilize impulse, and lead to guilt and debt.

Similarly, a developer in the gaming industry discussed extending users' time on platforms without considering addiction issues:

> The objective is to maximize player retention, increase usage time—that's what determines the company's market value. Our team had designers, psychologists—all working to retain as much time as possible, exploiting every trigger. It's our job. But will this affect my child, their generation?

These reflections on negative effects on users extend to the development of work platforms. A developer working on a delivery platform recounted suggesting that the system consider couriers' physical fatigue, transportation mode, and working hours. The suggestion was ignored:

> A delivery route optimization algorithm works with the classic 'traveling salesman problem.' But what variables are we taught to consider? The courier's body isn't one of them. It doesn't occur to most of my colleagues that it could be different; nobody asks if the traveling salesman gets tired.

The lack of corporate responsibility on the social dimensions of their systems extends to data governance. Lucas, a cybersecurity specialist, voiced concerns about the commodification of personal data and erosion of privacy:

> Our systems often focus on cost-efficiency, but we must also imagine the long-term societal costs. There's a lack of accountability in the volume of data captured, much of which doesn't directly serve the end activity, and the fact that it continues to be stored, creating ever-larger data assets as companies merge. If not with today's CEO, there's always the possibility of future misuse under a different board or political regime.

Concerns about surveillance are underlined by the historical fragility of Latin American democracies, with various dictatorial regimes of political persecution. These reflections also highlight the dual role of socio-technical imaginaries in both inspiring innovation and serving as cautionary tales.

There's also a debate about people's rights to the data they produce. One worker pointed out that users of proprietary platforms currently lack the option to share their data with public or citizen initiatives. For example, during the COVID-19 pandemic, there was a demand for users of biometric devices to share health data with the Unified Health System (SUS), but legal and technical barriers of proprietary hardware and software prevented this (Seto, 2024a).

This debate about data ownership and sovereignty is part of a larger concern of workers with academic and activist backgrounds, or/and those in public companies about technological sovereignty. Pedro, an activist from a tech workers' group associated with an homeless movement, emphasized: "We need public funding to develop our own technological solutions, tailored to our cultural and societal needs. Relying on imported technologies means importing others' values and biases."

A specific challenge highlighted was the linguistic and in large language models (LLMs). Tools like semantic content analysis and computer vision, developed primarily in the Global North, often underperform in Latin American contexts. This underscores the need for region-specific development trained on local languages like Portuguese, local images and geospatial databases. However, workers also acknowledged structural challenges, such as limited domestic investment and brain drain. João, a data scientist at a Brazilian state-owned company, pointed out:

> I have access to a national supercomputer that costs $36 million, has 2.4 petaflops, and is entirely dedicated to AI development. However, it's a small fraction of what's needed to develop something like ChatGPT, not to mention the volume of training data, which is the main difference.

The technological dependence on Global North platforms also reinforces surveillance concerns about national security risks, reflecting a geopolitical dimension of the workers' socio-technical imagination and echoing nationalist and developmentalist ideas. Matias, an employee at Petrobrás, explained:

> We've implemented an internal interface and a specific legal contract with OpenAI because their general terms allow them to use our data for training. However, our company uses Microsoft Teams for internal communications. In other words, information of national security interest—where our oil reserves are, technical details about our AI patents—is circulating unencrypted on a platform from a foreign country that can pass information to its state.

Most professionals agreed that technological investment in Brazil is concentrated in sectors like oil and gas, agribusiness, mining, and finance. Contrary to expectations of associating digital technologies and AI with a knowledge-based economy, this focus links AI development to the maintenance of extractivism and highly polluting activities as the center of the country's economic insertion into the global market. There's also reflections about digital technologies reinforcing not only Brazil's predatory economic model but also existing social inequalities. Daniela, a data scientist, noted:

> Automated decision-making by companies and governments could exacerbate marginalization, particularly for populations in Latin America that don't generate data. There's a segment of the Brazilian population whose digital integration isn't of market interest—due to generational, geographical, or class reasons. Investments focus on creating models that serve consumer populations in wealthy urban centers, leaving others invisible to these systems. Therefore, biases emerge from the low representation of disconnected people in datasets, like people from the northeast [Brazil's poorest region].

Therefore, the same Global North's "data universalism" that renders the specific needs of the Global South invisible in the dominant data regime (Milan & Treré, 2019) is reproduced in local tech development, amplifying internal regional inequalities within the country. The relevance of center-periphery relations at various scales is also reflected in the worker's view that Brazil doesn't have strength on the geopolitical stage to actually realize its digital sovereignty, as Matias, a senior data scientist, puts it:

> Why should I read Brazilian law? It doesn't matter. I read the U.S. presidential act on AI regulation because this really shapes our future. Essentially, it's a policy to protect the U.S.'s comparative advantages over the rest of the world. The White House now regulates any sale of GPUs. For any AI model on a public cloud server, the U.S. gains the right to access any information. The U.S. is saying: we'll control, as much as possible, the global development of AI.

This perspective highlights the challenges faced by countries like Brazil in asserting control over their own technological futures within the broader sociotechnical imaginaries of their workers. The very concept of sovereignty is controversial and disputed in the workers' imagination, as this speech of Mario, a developer and activist expresses:

> We see a lot of academic writers concerned with state sovereignty in a discourse of national development. I'm more concerned with how the interests of the majority of the population will prevail over these technologies, regardless of which side of the border.

This vision reflects an imaginary of power over technology that goes beyond traditional state sovereignty and focuses more on the technological autonomy of local communities to shape technology in ways that serve their interests and promote social transformation. This idea is materialized in collective initiatives within the tech community that aim to align the development of digital technologies with social needs and to build alternative tech futures, specially for urban peripheries. One such initiative is the Technology Sector of the Homeless Workers Movement (MTST). As one of its members, Amaranto, describes it:

> The Technology Sector of the MTST is a collective of social activists who develop digital solutions that make people's lives easier, because today access is concentrated among the rich and middle classes in urban centers. In short, we are a tool for social struggle, using technology to empower our quebradas and favelas [Brazilian terms for poor urban peripheries and slums].

The Sector is actively developing platforms tailored to marginalized groups. They collaborate with Señoritas Courier, a collective of cisgender women and trans individuals providing bicycle delivery services in São Paulo. Together, they are creating an internal platform to organize orders and distribute tasks among workers. This collaborative process includes weekly planning meetings and technopolitical training, ensuring that the workers not only use the platform but also understand and contribute to its development.

Another activist project involves developing a mobile application for women from traditional extractivist communities to help organize their cooperative production and sell of products derived from the babaçu coconut. Developers faced challenges like designing for users who are illiterate. By creating an interface based on visual icons, audiovisual resources and ensuring the app functions offline on low-cost devices, they worked to guarantee technological autonomy for the users. It was also necessary to try to translate technical concepts with analogies to traditional knowledge and worldview of this community, as Carlos, a developer of the project, explains:

> How do I explain what an internet cookie is? It's a vigilance mechanism, but it also facilitates online operation. We explain that it's like a gossipy neighbor, we have to be careful what we say to her, but her knowledge of everyone's lives helps in the day-to-day running of the community.

This is an example of how tech workers' socio-technical imaginaries are constructed in convergences but also tensions between users with very different repertoires of relationships with digital technologies. As another worker pointed out, it is easy to overlook the fact that a mechanism designed to improve the user experience which works well on mobile phones from at least the last five years can become a technical barrier for someone still using a device from ten years ago - 'and those people do exist', they added.

However, technological innovation also raises optimistic expectations for overcoming social inequalities. One senior developer suggested that advances in AI's natural language processing could revolutionize interfaces, moving beyond screens and text input to audio-based interactions such as those with virtual assistants. This shift could have a significant impact on digital inclusion for illiterate populations. As long as the models understand the orality of these people, who tend to be older and use a very different language to the average internet user, and consider whether they can afford it, he reflects.

This challenge exemplifies the unique dimensions of addressing the relationship between technology and society in the Global South. Brazilian tech workers navigate a complex scenario, where their pursuit of technological innovation serves as both a means of

empowerment—through fostering technological sovereignty and addressing local needs—and a space where the boundaries between empowerment and dependency blur.

## Conclusion

Rather than drawing narrow conclusions, this chapter offers a broader perspective on the sociotechnical imaginaries of Brazilian tech workers. This exploratory research focuses on the lived experiences and reflections of the workers themselves, challenging traditional knowledge hierarchies by emphasizing their voices. While their perspectives are engaged with academic concepts such as algorithmic bias, digital sovereignty, and technological dependence, the goal is not to produce research solely for an academic audience. Instead, it deliberately avoids being driven by the author's hypotheses, theoretical conclusions, or literature-centered debates, prioritizing the workers' "thought categories" as the core of the analysis.

That said, I would like to offer a few final reflections on the results of the fieldwork. The emphasis on diversity, inclusion, and local innovation provides Brazilian tech workers with a sense of agency over their work. However, they also have the potential to perpetuate traditional economic models and social disparities, especially when their work is subordinated to the prevailing logic of extractive industries, precarious labor, and addictive platforms—ultimately reinforcing hegemonic power structures. This highlights the need to move beyond abstract debates on ethical principles on technology development to a focus on ethics in practice, emphasizing the context in which ethical decisions are made (Powell et al., 2022).

There is also a fundamental territorial dimension to brazilian tech worker's sociotechnical imaginaries, constantly reflecting unequal power relations between centers and peripheries in technological development—whether it's the influence of global powers on national development, regional inequalities within Brazil, or the social abysm between wealthy and poor neighborhoods within a city. The idea of technologies from the periphery and for the peripheral population is very important for many interviewees, especially those from the favelas.

Their imaginaries also express aspirations for solidarity with marginalized groups, including precarious platform workers, reinforcing findings from other countries (Irani, 2019). This solidarity ranges from challenging cisheteronormative values embedded in algorithmic filters to reflecting on how labor platforms should better serve the interests of their workers. This is a powerful dimension because it runs counter to the technosolutionist approach that reduces social reality to abstract mathematical optimization problems, reclaiming that embodied material life should be considered, as seen in the case of delivery workers' route optimization.

While there are controversies, there is growing awareness among Brazilian tech workers that algorithmic biases reflect not only the datasets but also the subjectivities of the developers themselves, linking the importance of diversity, equity, and inclusion (DEI) initiatives to tangible impacts on technological development. These findings contrast with dominant perspectives in Global North tech worker communities, where biases in models are primarily attributed to flawed training data, and DEI efforts are often seen as symbolic rather than integral to addressing the ethical challenges posed by AI (Browne et al., 2024).

Overall, the workers interviewed demonstrate deep reflections on the societal impact of the technologies they develop, particularly regarding social inequalities—whether amplifying them or attempting to contribute to their resolution. As a qualitative study, it is not possible to determine whether this is a dominant characteristic within the Brazilian tech worker community, and future quantitative research could aim to map these debates more broadly.

Future research in this area must also address the limited research on tech workers in Latin America and even less in Africa. A key area to explore is the new wave of unionization among tech workers, whose research is still concentrated in the global North and Asia. An analysis of similar movements in Latin America and Africa would reveal how they are shaping local workers' imaginations. Another understudied area is tech workers who become social influencers: while many focus on career issues, they are also driving important debates about the social impact of technology, particularly in relation to AI. In addition, research is needed on tech workers' perspectives and their ability to influence public regulation of Big Tech and AI.

Nevertheless, our findings underscore the importance of incorporating tech workers' visions more fully into the debate on technology development in the Global South, while also highlighting the need to discuss tech workers' research beyond Silicon Valley. In this sense, the perspectives of Brazilian tech workers are relevant to researchers, policymakers, industry leaders, and civil society around the world as they navigate the complex interplay between technology and society.

Kenzo Soares Seto (2025).Navigating the Sociotechnical Imaginaries of Brazilian Tech Workers. In Ergin Bulut, Julie Chen, Rafael Grohmann, and Kylie Jarrett (eds.). SAGE Handbook of Digital Labour. SAGE: 314-324.

**Kenzo Soares Seto (2025).Navigating the Sociotechnical Imaginaries of Brazilian Tech Workers. In Ergin Bulut, Julie Chen, Rafael Grohmann, and Kylie Jarrett (eds.). SAGE Handbook of Digital Labour. SAGE: 314-324.**